# Fourier Magnetic Imaging with Nanoscale Resolution and Compressed Sensing Speed-up using Electronic Spins in Diamond

K. Arai[1] *, C. Belthangady[2,3]*, H. Zhang[2,3]*, N. Bar-Gill[2,3,+], S. J. DeVience[4],

P. Cappellaro[5], A. Yacoby[3] and R. L. Walsworth[2,3,6,‡]

[1] Department of Physics, Massachusetts Institute of Technology, Cambridge, Massachusetts 02139, USA.
[2] Harvard-Smithsonian Center for Astrophysics, Cambridge, Massachusetts 02138, USA.
[3] Department of Physics, Harvard University, Cambridge, Massachusetts 02138, USA.
[4] Department of Chemistry and Chemical Biology, Harvard University, Cambridge, Massachusetts 02138, USA.
[5] Nuclear Science and Engineering Department, Massachusetts Institute of Technology, Cambridge, Massachusetts 02139, USA.
[6] Center for Brain Science, Harvard University, Cambridge, Massachusetts 02138, USA.
[+]Current address: Department of Applied Physics and Racah Institute of Physics, The Hebrew University of Jerusalem, Jerusalem 91904, Israel.
*These authors contributed equally to this work.
‡Corresponding author (rwalsworth@cfa.harvard.edu)

**Optically-detected magnetic resonance using Nitrogen Vacancy (NV) color centres in diamond is a leading modality for nanoscale magnetic field imaging,[1-3] as it provides single electron spin sensitivity,[4] three-dimensional resolution better than 1 nm,[5] and applicability to a wide range of physical[6-8] and biological[9] samples under ambient conditions. To date, however, NV-diamond magnetic imaging has been performed using "real space" techniques, which are either limited by optical diffraction to ≈250 nm resolution[10] or require slow, point-by-point scanning for nanoscale resolution, e.g., using an atomic force microscope,[11] magnetic tip,[5] or super-resolution optical imaging.[12] Here we introduce an alternative technique of Fourier magnetic imaging using NV-diamond. In analogy with conventional magnetic resonance imaging (MRI), we employ pulsed magnetic field gradients to phase-encode spatial information on NV electronic spins in wavenumber or "k-space"[13] followed by a fast Fourier transform to yield real-space images with nanoscale resolution, wide field-of-view (FOV), and compressed sensing speed-up.**

Key advantages of NV-diamond Fourier magnetic imaging, relative to real-space imaging, are: (i) spatially multiplexed detection,[14] which enhances the signal-to-noise ratio (SNR) for typical NV centre densities; (ii) a high data acquisition rate that can be further boosted with compressed sensing;[15,16] and (iii) simultaneous acquisition of signal from all NV centres in the FOV, which allows probing of temporally correlated dynamics and provides isolation from system drift. As described below, we apply the Fourier technique using a relatively simple apparatus, and demonstrate one-dimensional imaging of individual NV centres with <5 nm resolution; two-dimensional imaging of multiple NV centres with ≈30 nm resolution; and two-dimensional imaging of nanoscale magnetic field patterns, with magnetic gradient sensitivity ~14 nT/nm/Hz$^{1/2}$ and spatial dynamic range (FOV/resolution) ~500. We also show that compressed sensing can accelerate the image acquisition time by an order-of-magnitude using sparse sampling followed by convex-optimization-based signal recovery. With further improvements, NV Fourier magnetic imaging may enable MRI with atom-scale resolution and centimeter FOV, with applications ranging from mapping the structure of individual biomolecules to functional MRI in living cells to studies of quantum phenomena in magnetic materials. Similar Fourier techniques could also be applied to NV-diamond imaging of nanoscale electric fields, temperature, and pressure, as well as to other optically-addressable solid-state spin systems.

Schematic views of the Fourier magnetic microscope are shown in Figs. 1a and 1b. The diamond sample has a thin layer of NV centres at the surface created via ion implantation (see Methods). NV electronic spin states (Fig. 1c) are optically polarized with green illumination (λ=532 nm), coherently manipulated using resonant microwave fields applied by a microwave loop, and detected via spin-state-dependent fluorescence measurements. Strong, uniform magnetic field gradients for NV spin phase-encoding are generated by currents sent though pairs of wires (gradient microcoils) separated by 100 μm and connected in an anti-Helmholtz configuration. Another current sent through an external-field wire produces a non-uniform magnetic field pattern that can be imaged with the ensemble of NV sensors in the sample. The microwave loop, gradient microcoils, and external-field wire are patterned using photolithography onto a diamond coverslip, which is bonded to the NV-containing diamond sample with optical adhesive.

Fig. 1b displays a simulation of the gradient produced by a current of 1 A sent through the gradient microcoils. At the centre of the microcoils, the gradient magnitude $|dB_\zeta/dx| = |dB_\zeta/dy| \approx 0.7$ G/μm and varies by less than 1% over a 15 μm x 15 μm region. Here $\zeta$ represents the NV quantization axis as shown in Fig. 1a.

The Fourier magnetic imaging pulse sequence consists of a laser initialization pulse, a microwave dynamical-decoupling sequence for spin-state manipulation,[17,18] and a laser readout pulse (Fig. 1d). A pulsed gradient field ($dB_\zeta/dx$ and/or $dB_\zeta/dy$) for NV spin phase-encoding is applied during the microwave sequence. The sign of the gradient can be switched during alternate free precession intervals to enable AC magnetic field sensing. The NV spin coherence thus acquires a position-dependent phase $\varphi = 2\pi \vec{k} \cdot \vec{r}_0$, where $\vec{k} = \gamma\tau(dB_\zeta/dx, dB_\zeta/dy)$ defines the imaged point in the two-dimensional Fourier or k-space, and $\vec{r}_0$ denotes the position of the NV. Here $\gamma$ = 2.8 MHz/Gauss is the NV gyromagnetic ratio and $\tau$ is the total precession time. An optical readout pulse measures the NV signal for this point in k-space, which is proportional to the cosine of the acquired phase $\varphi$: i.e., $s(\vec{k}) \sim \cos(2\pi \vec{k} \cdot \vec{r}_0)$. By incrementally stepping through a range of amplitudes for the applied magnetic field gradient, while keeping the precession time $\tau$ fixed, the NV signal is measured as a function of $\vec{k}$ to produce a k-space image. The real-space image is then reconstructed by Fourier transformation of the k-space image: i.e., $S(\vec{r}) = F[s(\vec{k})]$ where $Abs[S(\vec{r})]$ gives the positions of the NV centres. The pixel resolution of the real-space image is $(2k_{max})^{-1}$, where $k_{max} = \gamma\tau(|dB_\zeta/dx|, |dB_\zeta/dy|)_{max}$ is the maximum k value used in the measurement.

As a first demonstration of NV Fourier imaging, we acquired a one-dimensional image of a single NV centre in a low NV density sample (Sample A, see Methods) by varying only the $dB_\zeta/dx$ gradient strength and using a 4-pulse Carr-Purcell-Meiboom-Gill (CPMG) decoupling sequence. The image was acquired for k values between 0 and 0.144 nm$^{-1}$, with clear oscillations with a period of 0.0025 nm$^{-1}$ seen in the k-space image (Fig. 2a). Upon Fourier transformation, the k-space image produced a real-space image with a single peak indicating the single NV location. The peak width and amplitude correspond to a real-space pixel resolution of 3.5 nm and SNR=13 (Fig. 2b). This resolution surpasses the highest demonstrated to date for NV centres using STED super-resolution microscopy (≈6 nm).[19] The data acquisition time per point in k-space was $T_{DAQ} = T/(N_{pix} * SNR^2)$ ≈20 ms, where $T$ is the total imaging time and $N_{pix}$ is the number of pixels in the real-space image. We observed no significant broadening of the real-space signal peak at this image resolution, indicating that there was insignificant effect from technical issues such as gradient current instability, thermal fluctuations, mechanical vibrations, and laser instability.

We next demonstrated two-dimensional Fourier imaging of multiple NV centres using a high-NV-density sample (Sample B, see Methods). To create a regular pattern of NV centres to be imaged for this and later demonstrations, we fabricated arrays of nanopillars (400 nm diameter, 100 nm height, 1 μm spacing) on the diamond surface. We chose the implant dosage and nanopillar size such that there were, on average, two NV centres of the same crystallographic orientation per nanopillar. Scanning electron microscope (SEM) and confocal microscope images of a typical nanopillar, close to the centre of the gradient microcoils, are shown in Figs. 2c and 2d, respectively. We used a spin-echo sequence and varied both the $dB_\zeta/dx$ and $dB_\zeta/dy$ gradient strengths to acquire a two-dimensional k-space image of the NV centres in this nanopillar, as shown in Fig. 2e. A Fourier transform of the k-space image, thresholded at five-sigma above the noise level (see Supplementary Information), reveals two proximal NV centres separated by 121(9) nm with a pixel resolution of 30 nm (Fig. 2f). This resolution is consistent with the k-space spanned in Fig. 2e. For this two-dimensional Fourier image, the data acquisition time per point in k-space was $T_{DAQ}$ ≈15 ms.

Once the locations of individual NV centres are precisely determined, they can be used as high sensitivity probes to measure magnetic fields that vary on length scales smaller than the optical diffraction limit. To demonstrate such nanoscale Fourier magnetic imaging, we introduced a spatially inhomogeneous magnetic field over the FOV in the centre of the gradient microcoils by sending an alternating electric current, phase-locked to the dynamical-decoupling microwave pulse sequence (see Fig. 1d), through the external-field wire shown in Fig. 1a,b. NV centres within this region were thus exposed to different AC magnetic field amplitudes depending on their locations relative to the external-field wire. For a fixed value of the alternating current, the k-space signal for a given NV at location $\vec{r}_0$ acquires a phase offset $\theta = 2\pi\gamma B_{ext}\tau$ where $B_{ext}$ is the AC magnetic field amplitude (projected along the NV axis) at the site of the NV: i.e., $s(\vec{k}) \sim \cos(2\pi\vec{k} \cdot \vec{r}_0 + \theta)$. As before, Fourier transforming the k-space image produces a real-space image with $Abs[S(\vec{r})]$ showing a peak at $\vec{r} = \vec{r}_0$. In addition, the argument $Arg[S(\vec{r}_0)]$ yields the phase shift $\theta$ from which $B_{ext}$ can be determined. In Fig. 3 we illustrate how measured values for $B_{ext}$ can be uniquely assigned to the two NV centres shown in Fig. 2f. For each value of the applied alternating current, we recorded a one-dimensional k-space image by incrementally varying the $dB_\zeta/dx$ gradient strength. Fig. 3a shows the resulting values of $Abs[S(x)]$ as a function of current through the external-field wire with a threshold set at five-sigma (see Supplementary Information). We consistently observe two distinct peaks (corresponding to the two NV locations) through the entire current range. The cosine of $Arg[S(x_i)]$, where $x_{i=1,2}$ denotes the location of each peak, is shown in Figs. 3b,c. When the applied current is very small (i.e., weak $B_{ext}$ gradient), the two NV centres are exposed to nearly the same value of $B_{ext}$ and consequently there is little difference in the measured NV k-space phases (left panel of Fig 3c). However, for a relatively large current of 5.13 mA (i.e., stronger $B_{ext}$ gradient, right panel of Fig 3c), the measured NV k-space phase difference is $2\pi \times (2.6 \pm 0.4) \times 10^{-1}$ radians, which yields a magnetic field difference of $\Delta B_{ext} = (6.5 \pm 1.1) \times 10^2$ nT between these two NV centres separated by ≈121 nm, equivalent to a magnetic gradient sensitivity of ~14 nT/nm/Hz$^{1/2}$ (see Supplementary Information). The measured field and gradient values are consistent with those expected from simulations.

NV-diamond Fourier magnetic imaging can also be integrated with wide-field, real-space microscopy to realize a very large spatial dynamic range: i.e., magnetic imaging with both wide FOV and nanoscale resolution. As a demonstration of such hybrid real + k-space imaging, we scanned the Fourier magnetic microscope across a wide array of NV-containing nanopillars in Sample B and performed both low-resolution real-space and high-resolution k-space NV imaging of magnetic field patterns produced by the external-field wire. Real-space image resolution is limited by optical diffraction, while nanometer-scale information is obtained via Fourier magnetic imaging. For example, we sent an AC current of 5.13 mA and frequency 50 kHz through the external-field wire and performed low-resolution real-space magnetic imaging on 167 nanopillars in a 15 µm x 15 µm region near the edge of the wire (Fig. 4). From these measurements, we estimated the magnitude of the magnetic field gradient at each nanopillar location. Next, we performed high-resolution k-space imaging (followed by Fourier transforming into real-space) of both the location of individual NV centres in several nanopillars with 30 nm spatial resolution, as well as the magnetic field difference $\Delta B_{ext}$ sensed by the NV centres within each nanopillar. As shown in Fig. 4, the resulting nanoscale maps of $\Delta B_{ext}$ are consistent with the local magnetic field gradient values obtained with low-resolution real-space imaging. Together, this hybrid real + k-space imaging demonstration provides a spatial dynamic range (FOV/resolution) ≈500. In future work we plan to integrate NV Fourier magnetic imaging with a wide-field real-space imager (e.g., using a CMOS or CCD camera as in refs. 9 and 10), which would obviate the need for scanning and provide parallel and hence rapid real-space imaging across a very wide FOV >1 mm[20] together with nanoscale resolution k-space imaging. The key technical challenge for this approach will be achieving strong, uniform pulsed magnetic field gradients across such a large FOV.

Finally, we showed that the speed of NV Fourier magnetic imaging can be greatly enhanced using compressed sensing techniques.[15,16] In compressed sensing, signals are sampled at random and at a sampling frequency that may be lower than the Nyquist rate. Under assumptions of signal sparsity, recovery algorithms based on convex-optimization can then be used to reconstruct the signal in a transform domain with high fidelity. We first recorded a fully sampled (N = 2048 data points), one-dimensional, k-space image of two NV centres in Sample B by sweeping the $dB_\zeta/dx$ gradient (blue trace of Fig. 5a). In subsequent measurements we recorded data at $M = N/2^p$ (*p* = 0,…,6) randomly chosen k-space values, providing a factor of *N/M* speed-up (under-sampling) in data acquisition. For example, the red dots in Fig. 5a show data collected for *M* = 128 k-space values and hence a speed-up of *N/M* = 16. A Fourier transform of the fully sampled curve shows two NV centres separated by 116 (4) nm (blue trace in Fig. 5b). To reconstruct real-space images from the under-sampled data sets, we minimized the $l_1$-norm of the real-space signal $\|\vec{x}\|_1$ subject to the constraint $\vec{k} = \overleftrightarrow{A} \cdot \vec{x}$, where $\vec{x}$ and $\vec{k}$ are column vectors of size $N$ and $M$ representing real-space and k-space data respectively, and $\overleftrightarrow{A}$ is a Fourier-transform sampling matrix of size $M \times N$. The red traces in Fig. 5b show the absolute values of reconstructed real-space signals (offset for clarity). For $M \geq 128$ we observe two peak positions that match well with the NV locations given by the fully-sampled data; this result is consistent with the criterion for faithful signal recovery in compressed sensing, which requires $M \geq CS\text{Log}_2(N)$, where $S$ is the signal sparsity (equal to 2 in the example above) and $C$ is a constant of order 1, which in this case is found to be ≈5. Importantly, we also demonstrated that the real-space signal reconstructed via compressed sensing retains reliable information about the magnetic field sensed by each NV centre. We sent an AC current of 10.26 mA and 50 kHz through the external-field wire, recorded k-space data for $M = 128$ points, and reconstructed real-space signals via compressed sensing as described above. From the phase of the reconstructed signals we determined the difference between the magnetic fields sensed by the two NV centres to be $(5.0 \pm 0.5) \times 10^2$ nT, in reasonable agreement with the value of $(6.3 \pm 0.8) \times 10^2$ nT obtained using fully-sampled k-space data, but with the benefit of a compressed sensing speed-up factor of 16. The reliable retention of NV phase information in data reconstructed via compressed sensing can be made explicit by computing the inverse Fourier transform of the reconstructed signal. In Fig. 5c we compare NV k-space signals obtained via such inverse Fourier transformations *with* (black trace) and *without* (red trace) current in the external-field wire. The observed phase shift between the two traces provides a direct illustration and measure of the magnetic field difference between the two NV centres, consistent with the above results.

Our demonstration of Fourier magnetic imaging provides the first method for mapping NV positions and local magnetic fields in wavenumber or "k-space," which can then be Fourier transformed to yield real-space images with both nanoscale resolution and wide field-of-view (FOV). The distinct advantage of this approach, relative to real-space imaging, is *spatially multiplexed signal acquisition across the full FOV*, which enhances the SNR for typical NV centre densities; enables a higher data acquisition rate that can be increased by more than an order-of-magnitude with compressed sensing; and allows one to probe classical or quantum correlations in samples by conducting simultaneous measurements using multiple, spatially-separated NV centres. We also emphasize the relative simplicity of the apparatus needed for Fourier imaging, with the micro-gradient coils used for phase-encoding being easily integrated with an optical microscope. We expect that NV Fourier magnetic imaging will be applicable to a broad range of systems that can be placed on or near the diamond surface. Example applications in the physical sciences include probing quantum effects in advanced materials, such as frustrated magnetic systems with skyrmionic ordering; spin liquids where quantum spin fluctuations prevent the system from ordering; low-dimensional systems such as graphene, as well as anti-ferromagnetic and multiferroic materials; and topological insulators with quantized spin-carrying surface states. In the life sciences, NV Fourier magnetic imaging may allow nanoscale NMR spectroscopy[21] and structure determination of individual biomolecules; MRI within living cells; and real-time, noninvasive mapping of functional activity in neuronal networks with synapse-scale resolution (~10 nm) and circuit-scale FOV (>1 mm). Anticipated technical improvements of NV Fourier imaging include: (i) enhanced magnetic field sensitivity via optimization of diamond samples,[22] optical collection efficiency,[23] and spin-state optical contrast; (ii) two-dimensional imaging resolution <1 nm, made possible by stronger magnetic field gradients generated by smaller microcoils and the use of dynamical decoupling pulse sequences to extend NV spin coherence times;[18,24,25] and (iii) parallel real-space image acquisition with a wide-field CMOS or CCD camera, which will provide immediate improvement in the spatial dynamic range (FOV/resolution) and enable study of long-range, real-time dynamics such as neuronal activity that span length scales from a few nanometers to millimeters. Straightforward extensions of the present technique should also allow NV Fourier imaging of electric fields,[26] temperature,[27] and pressure[28] with nanoscale resolution, wide FOV, and compressed sensing speed-up, as well as application to other solid-state quantum spin systems such as point defects in silicon carbide.[29]

## Methods

### NV physics

The ground state of the NV centre is a spin-triplet with a 2.87 GHz zero-field splitting between the $|0\rangle$ and $|\pm 1\rangle$ spin states (Fig. 1c). An applied magnetic field defines a quantization axis and Zeeman splits the $|\pm 1\rangle$ states. Under optical excitation at 532 nm, the NV centre undergoes spin-state-preserving transitions between the electronic ground and excited states, emitting fluorescence in the 640 nm to 800 nm band. A non-radiative decay pathway from the $|\pm 1\rangle$ excited states to the $|0\rangle$ ground state allows optical initialization of the NV centre, as well as optical readout of the spin state via spin-state-dependent fluorescence intensity.

### Fabrication of nanopillars and gradient coils

Samples A and B were electronic-grade, single-crystal diamond chips with natural isotopic concentration of $^{13}C$ (1.1%), which were grown using chemical vapour deposition (Element 6 Corporation). Sample A was implanted with $^{14}N$ ions (dosage $10^{10}$ cm$^{-2}$; energy 85 keV). From SRIM simulations the NV centres were estimated to be 100 nm below the diamond surface. Sample B was implanted with $^{15}N$ ions (dosage $10^{12}$ cm$^{-2}$; energy 14 keV) and the estimated NV depth was 20 nm. The conversion efficiency from implanted N ions to NV centres in both samples was estimated to be ~1%. Typical NV spin dephasing times were $T_2^* \approx 1$ μs (Sample A) and 500 ns (Sample B); while the typical Hahn-echo spin coherence times were $T_2 \approx 50$ μs (Sample A) and 30 μs (Sample B). Magnetic field gradients used in the Fourier imaging demonstrations did not significantly affect NV spin coherence properties across the imaging FOV.[30] To fabricate diamond nanopillars, Sample B was spin coated with a 100 nm-thick layer of e-beam resist (XR-1541-006). Arrays of 400-nm-diameter circles were then patterned using an Elionix ELS-7000 e-beam writing system with exposure dosage and beam energy set at 8,000 μC/cm$^2$ and 100 kV respectively. Tetra-methyl ammonium hydroxide (25%) was used to develop the resist and form the etch mask. The diamond crystal was then placed in an inductively coupled plasma (ICP) reactive-ion etching system and etched for 45 s with 30 s.c.c.m. of oxygen gas, 700 W ICP power, and 100 W bias power at a chamber pressure of 10 mtorr.

These parameters gave an etch depth of ~150 nm. The gradient coils were fabricated on a polycrystalline diamond coverslip (10 mm × 10 mm × 0.3 mm) for optimized heat dissipation. The coverslip was spin coated with LOR 20B (bottom layer ~2 µm) and Shipley S1805 (top layer ~400 nm) photoresists, exposed using a mask aligner, and the pattern was developed in a CD26 bath. A 30 nm Ti layer and a 970 nm Au layer were then deposited in an e-beam evaporator, followed by lift-off in MicroChem Remover PG solution. Electrical resistance measurements, performed using a four-probe station, gave a value of ~1 Ω for each gradient coil. A heat sink was attached to the back surface of the diamond coverslip to enhance heat dissipation.

**Data analysis**

Fourier magnetic imaging signals were recorded in k-space using the techniques described in the main text. A tapered-cosine windowing function with taper coefficient set at 0.1 was applied to the k-space data and a symmetric fast Fourier transform algorithm implemented with MATLAB was used to obtain real-space images. No zero padding was done on the k-space data. Real-space pixel resolution therefore matched the true resolution of the imaging system. The SNR of real-space images acquired with this method increased with the square root of the number of data points, which confirmed the multiplex advantage of the Fourier imaging technique (see Supplementary Information). The real-space images in Figs. 2, 3 and 4 were thresholded at $5\sigma$ above the noise level ($\sigma$ is the standard deviation of the noise); and NV centres were represented by circles centred at the positions of the thresholded peaks, with diameters equal to the pixel resolution. For the wide FOV image of Fig. 4, the k-space sampling rate and number of sample points were held fixed for all nanopillars. For the magnetic field sensing data of Fig. 4, a calibration procedure was first performed by recording NV fluorescence as a function of current through the external-field wire for each nanopillar in the FOV without phase encoding. The resultant magnetometry curves revealed the functional dependence of magnetic field on the applied current for each nanopillar. For a fixed current, the magnetic field so determined was used as an offset value for the high-spatial-resolution phase-encoded magnetic field data recorded from individual NV centres within the nanopillar of interest. Magnetic field sensitivity was calculated as $(\sigma_B/dS_B)T^{1/2}$, where $\sigma_B$ is the meas-urement noise, $dS_B$ is the slope of the magnetometry curve, and T is the total data acquisition time. For the NV centres and measurement protocols employed in this work, the magnetic field sensitivity was ~1 $\mu T/Hz^{1/2}$.

**Compressed sensing**

The under-sampled k-space data was windowed with a tapered-cosine function (taper parameter = 1). A sampling matrix, $\overleftrightarrow{A}$ of size $M \times N$ was created by picking $M$ rows (corresponding to the sampled points) from an $N \times N$ discrete Fourier transform matrix. The real-space signal was reconstructed by minimizing the $l_1$-norm of the real-space signal $\|\vec{x}\|_1$ subject to the constraint $\vec{k} = \overleftrightarrow{A} \cdot \vec{x}$. The convex optimization routine was implemented using MATLAB library functions made available by CVX Research (www.cvxr.com).

## Acknowledgements

This work was supported by the NSF, MURI QuISM, and DARPA QuASAR programs. We gratefully acknowledge the provision of diamond samples by Element 6 and helpful technical discussions with Mathieu Sarracanie, Matthew Rosen, David Phillips, Alex Glenday, and Birgit Haussmann.

## Author Contributions

K.A., C.B., and H.Z. contributed equally to this work. R.L.W. conceived the idea of NV Fourier magnetic imaging and supervised the project. K.A., C.B., and H.Z. developed measurement protocols, hardware, and software for NV Fourier magnetic imaging, performed the measurements, and analyzed the data. C.B. and N.B.-G. developed the NV-diamond confocal microscope used in the study. N.B.-G. also aided the development of data acquisition software. S.J.D. and A.Y. advised Fourier imaging techniques and applications. P.C. advised compressed sensing techniques and applications. All authors discussed the results and participated in writing the manuscript.

## Competing Financial Interests

The authors declare no competing financial interests.

**Figure legends**

**Figure 1 | Fourier magnetic imaging experiment. a**, Schematic of the Fourier magnetic imaging microscope. NV centre magnetic sensors are located near the surface of a diamond chip (e.g., as represented by spheres with arrows). NV spin states are initialized and read out with a green (532 nm) laser, and coherently manipulated with resonant pulses using a microwave loop. Controlled magnetic field gradients for NV spin phase-encoding are generated by sending currents through pairs of wires (gradient microcoils). An external-field wire is used to create a non-uniform DC or AC magnetic field for demonstrations of nanoscale Fourier magnetic imaging. The NV quantization axis, represented by $\zeta$, is offset from the surface normal (z-axis) of the [100]-cut diamond sample and aligned with a static, uniform magnetic field ≈30 G created by a permanent magnet (not shown). **b**, Top-view schematic of Fourier magnetic imaging microscope, along with a simulation (using COMSOL Multiphysics) of the magnetic field gradient $\sqrt{(dB_\zeta/dx)^2 + (dB_\zeta/dy)^2}$ when a current of 1 A is sent through both microcoil pairs, with current directions indicated by white arrows. Scale bar is 20 μm. **c**, Energy-level diagram of the NV centre; see Methods for details. **d**, Fourier magnetic imaging experimental sequence. Spins are polarized into $|0\rangle$ state with a green laser pulse. A microwave pulse sequence with $2n$ $\pi$ pulses dynamically decouples NV centres from magnetic field noise from the environment. A pulsed magnetic field gradient of alternating direction is applied during each precession interval. Spins at different locations accumulate phase at different rates. A final $\pi/2$ pulse projects the spins into the $|0\rangle$-$|1\rangle$ manifold; and state populations are read out optically via spin-dependent fluorescence. An AC magnetic field $B_{ext}$ produced by current in the external-field wire can be sensed using the procedure described in the main text.

**Figure 2 | Fourier imaging of NV centres with nanoscale resolution.** **a**, One-dimensional k-space image of a single NV centre in diamond (Sample A). As the gradient strength $dB_\zeta/dx$ is incrementally stepped by varying current through the microcoil, the NV fluorescence shows sinusoidal oscillations. Here we show NV fluorescence normalized to a reference measurement of $|0\rangle$ state fluorescence and with a constant background level subtracted. **b**, One-dimensional real-space image data (black) obtained from the absolute value of the Fourier transform of the k-space data. For comparison, the diffraction-limited, real-space point-spread-function of the microscope is shown (pink shaded area, full-width-half-maximum =300 nm). **c**, Scanning electron micrograph of a 400 nm diameter NV-containing diamond nanopillar fabricated on Sample B. **d**, Scanning confocal fluorescence image of the same nanopillar. Full width half maxima of corresponding x and y-profiles ≈ 400 nm. **e**, Two-dimensional k-space image of two proximal NV centres inside this same nanopillar. **f**, Fourier transformed, two-dimensional real-space image (absolute value) with a threshold set at 5σ above the noise level, where σ is the standard deviation of observed optical noise. Two NV centres separated by 121 nm are clearly resolved. Scale bars are 200 nm for **c**, **d**, and **f**, and 0.0066 nm$^{-1}$ for **e**.

**Figure 3 | Fourier magnetic gradient sensing below the optical diffraction limit.** One-dimensional k-space magnetic images for the two NV centres shown in Fig. 2f are acquired by incrementally stepping the $dB_\zeta/dx$ gradient strength for fixed values of AC current sent through the external-field wire. For each current value, the corresponding k-space image is Fourier transformed and thresholded at 5σ to obtain a one-dimensional real-space image. **a,** Absolute value of the real-space image shows peaks corresponding to the two NV centres separated by 121 nm. The vertical axis is real-space position along the x direction; and the horizontal axis corresponds to different ranges of AC currents at frequency 50 kHz, increasing from left to right: 0-0.51 mA, 2.56-3.08 mA, and 5.13-5.64 mA. **b**, Cosine of the argument of the real-space images of the two NV centres shown in **a**, for the same ranges of AC currents. **c**, Overlay of measured values for $\cos\theta$ (points) and corresponding fits (solids curves) as a function of AC current amplitude for the two NV centres, where $\theta = 2\pi\gamma B_{ext}\tau$. The observed differential phase shift between the data for the two NV centres shows that the these spatially-separated NV centres measure a magnetic field difference $\Delta B_{ext}$ arising from a gradient in the external AC magnetic field magnitude $B_{ext}$.

**Figure 4 | Fourier magnetic imaging with wide field-of-view and nanometer-scale resolution.** The AC magnetic field produced by passing a 50 kHz, 5.13 mA electric current through the external-field wire (indicated by a thick yellow line at the top-left corner) is imaged using a hybrid real + k-space technique over a wide field-of-view (FOV) spanning 167 diamond nanopillars. A low-resolution real-space magnetic image is acquired over the full FOV by scanning the microscope across all nanopillars (see Methods). The spatial resolution is limited by optical diffraction and NV centres within individual nanopillars are not resolved. Fourier (k-space) magnetic imaging is then performed on individual nanopillars to determine NV centre positions and local AC magnetic field amplitudes with ≈30 nm resolution (right boxes of inset panels). To check for consistency, the measured long-range magnetic field gradient provided by the low-resolution real-space image, together with the NV positions determined via Fourier imaging, are used to estimate the variation in AC magnetic field amplitude within each nanopillar (left boxes of inset panels). Good agreement is found between the measured and estimated values for the magnetic field difference between NV centres within each nanopillar (see Supplementary Information). Scale bar is 2 µm for main figure and 100 nm for inset panels.

**Figure 5 | Compressed sensing speed-up of NV Fourier magnetic imaging. a**, For two NV centres in Sample B: fully sampled one-dimensional k-space signal with $N = 2{,}048$ data points (blue line); and randomly under-sampled k-space signal with $M = 128$ data points (red dots). Under-sampling (speed-up) factor is $N/M = 16$. **b**. Blue trace shows absolute value of Fourier transform of fully sampled k-space data, indicating a real-space NV separation of 116 nm along the x-axis. Red traces (offset for clarity) show real-space signals reconstructed from under-sampled k-space data sets via compressed sensing techniques, in good agreement with fully sampled k-space data for $M \geq 128$ (see text for details). **c**, Inverse Fourier transform of data reconstructed via compressed sensing for $M = 128$ *with* (black trace) and *without* (red trace) an AC current (50 kHz, 10.26 mA) sent through external-field wire. The observed phase shift between the data sets provides a measure of the magnetic field difference between the positions of the two NV centres, in good agreement with the results from fully sampled k-space data, thereby showing that compressed sensing reconstruction retains reliable information about imaged magnetic fields.

**Figure 1:**

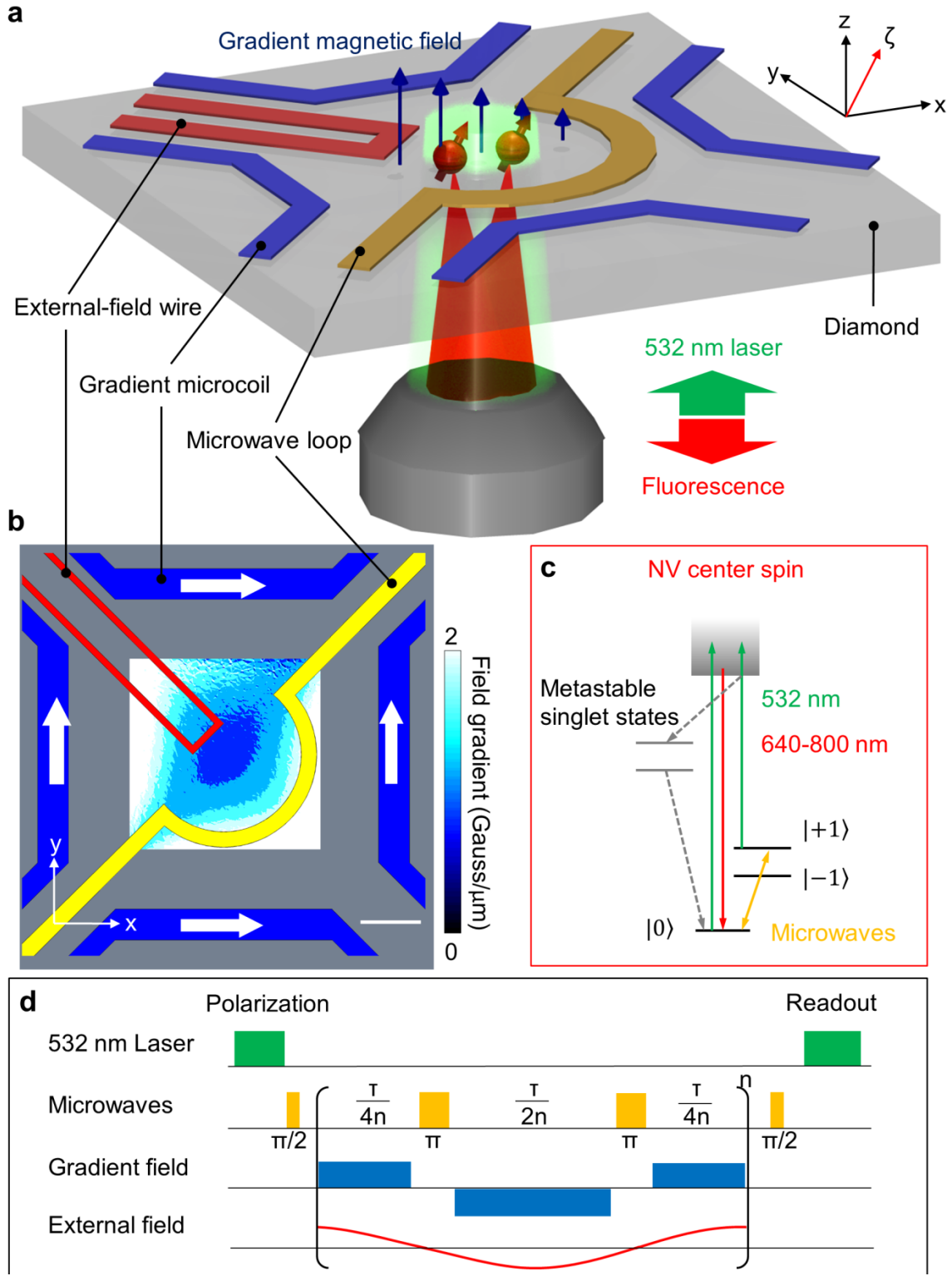

**Figure 2:**

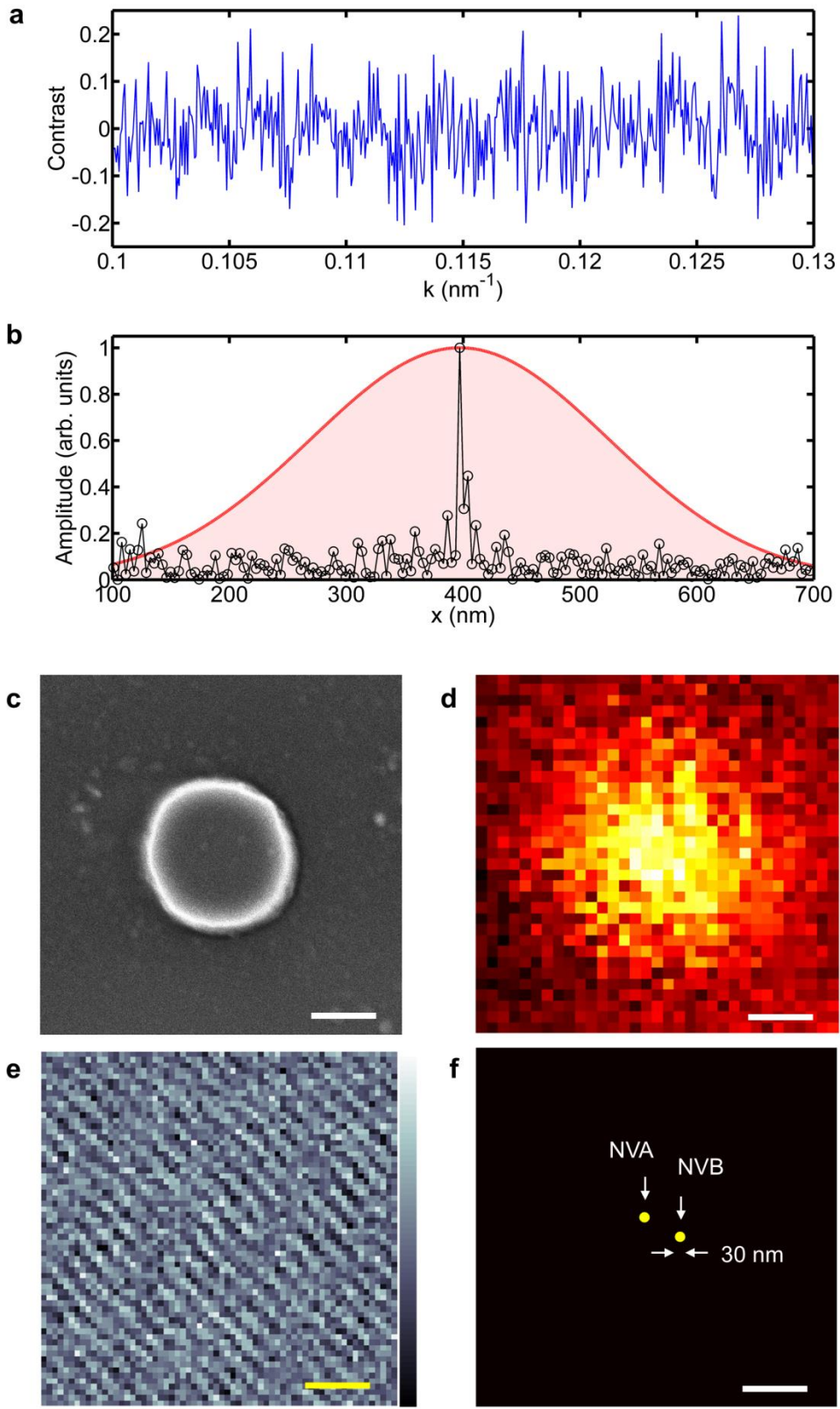

**Figure 3:**

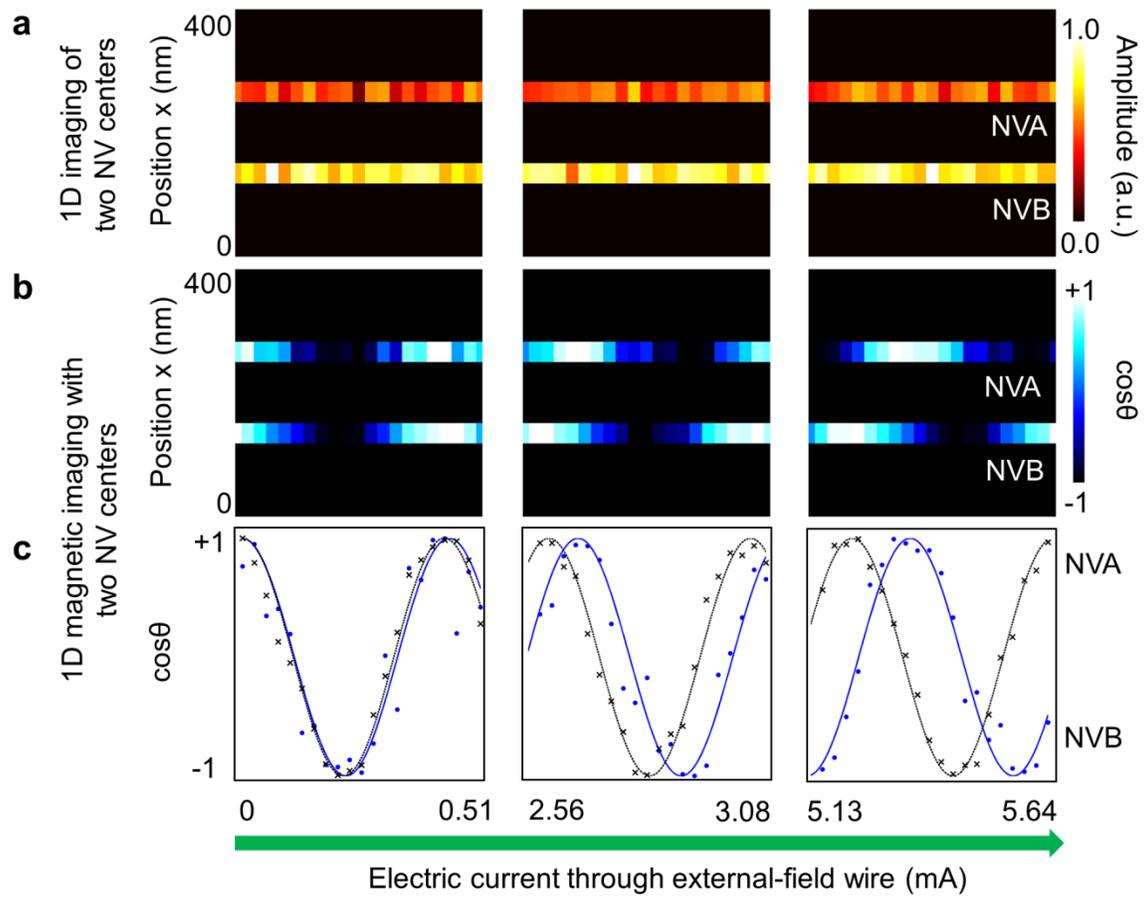

**Figure 4:**

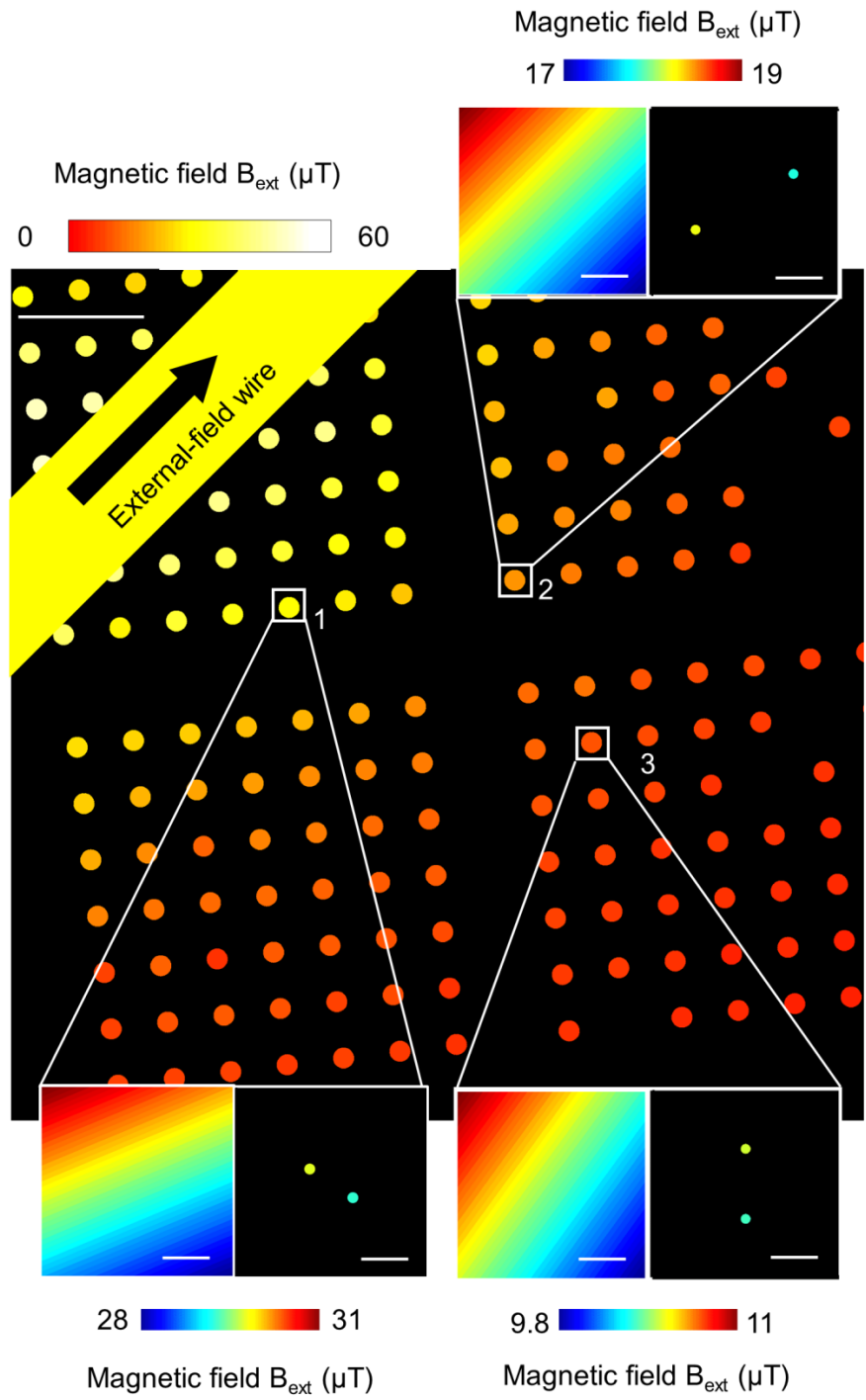

**Figure 5:**

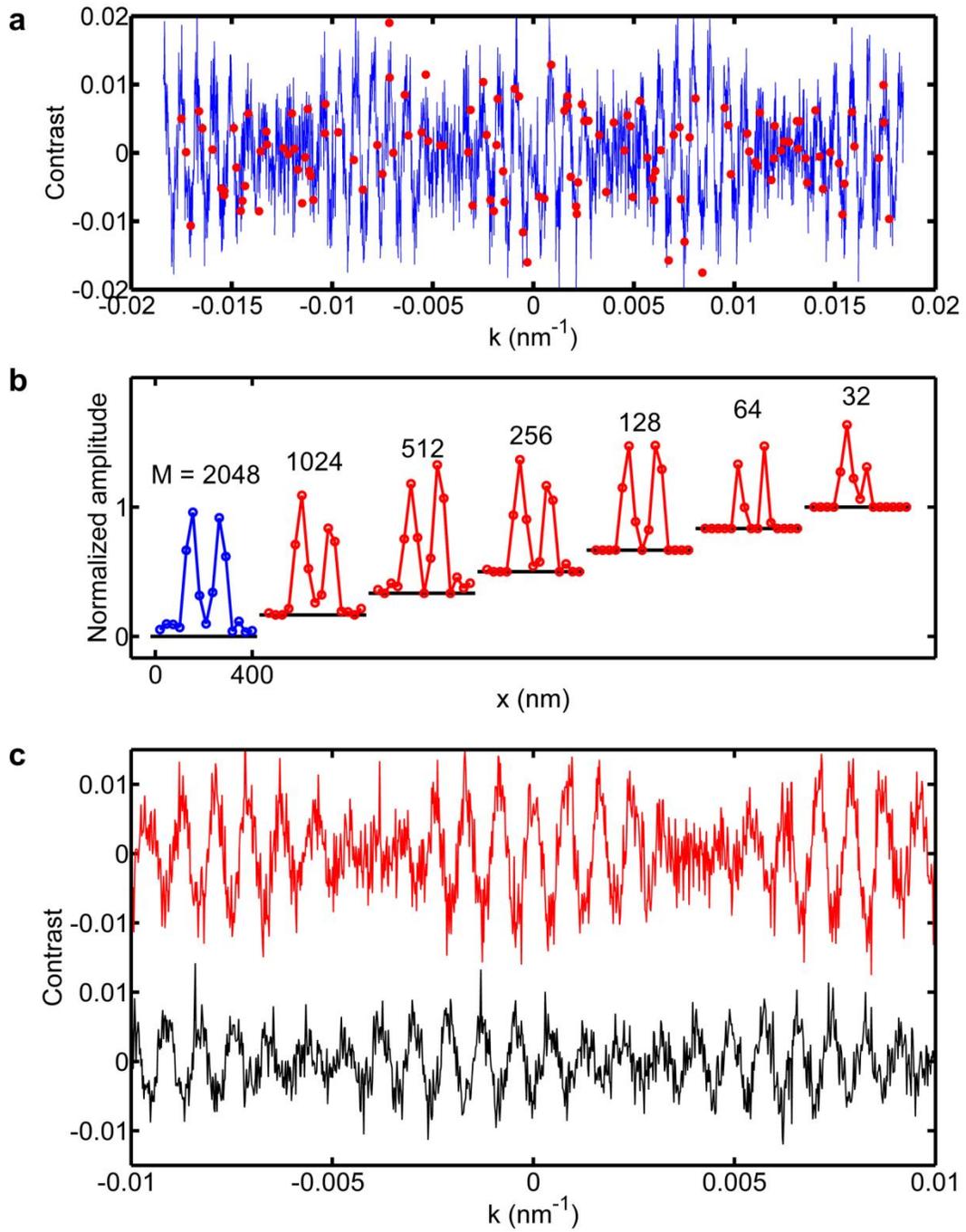

# Supplementary Material for "Fourier Magnetic Imaging with Nanoscale Resolution and Compressed Sensing Speed-up using Electronic Spins in Diamond"

## I. Fourier magnetic microscope

The Fourier magnetic microscope consisted of a home-built scanning confocal microscope and a diamond coverslip on which gradient microcoils were photo-lithographically defined. Optical excitation for the confocal microscope was provided by a 400 mW diode-pumped solid state laser (Changchun New Industries) operating at 532 nm. An 80 MHz acousto-optic modulator (Isomet Corporation) controlled laser pulses for spin initialization and read-out. A single-mode fiber was used to clean up the mode profile of the laser and the output of the fiber was sent to a 100X oil-immersion objective with a numerical aperture of 1.3 (Nikon CFI Plan Fluor). The NV-containing diamond sample was mounted on a three-axis motorized stage (Micos GmbH) that allowed scanning of the sample in the focal plane of the objective. Fluorescence from NV centres was separated from the 532 nm excitation light with a dichroic beam-splitter (Semrock LM01-552-25). After filtering through a long-pass filter (Semrock LP02-633RS-25), the fluorescence was focused into a single-mode fiber with a mode-field-diameter ~5 μm (which acted as the pinhole of the confocal microscope) and sent to a single-photon counting module (Perkin Elmer SPCM-ARQH-12). Pulses from the detector were read using the counter input of a DAQ card connected to a computer. Microwave excitation was provided by a signal generator (Agilent E4428C) whose output was amplified (Mini-circuits ZHL-16W-43-S+) and sent though a microwave loop defined on the diamond coverslip. Fabrication of the photo-lithographic patterns on the diamond coverslip is described in the Methods section of the main text. Gradient pulses were defined using the arbitrary waveform function of a programmable signal generator (Stanford Research Systems SRS345) and amplified by an audio-frequency amplifier (Texas Instruments LM4780). The bandwidth of the gradient pulses was limited by the slew rate of the amplifier to approximately 1 MHz.

## II. Thermal effects of gradient microcoil operation

The resistance of each gradient microcoil was 1.4 Ω, such that 1.4 W of heat was generated at a typical peak current of 1 A. A copper heat sink cooled by a fan was attached to the back surface of the diamond coverslip to promote effective heat dissipation. To assess thermal performance under standard operating conditions, the system was modeled using COMSOL Multiphysics software. With a heat sink and fan cooling we estimated a steady-state temperature rise of the diamond sample of <10 K at 1 A peak current. An analytical model based on techniques described in Ref. S1 gave similar temperature estimates.

The rise in temperature was experimentally determined using two techniques. In the first technique, electron spin resonance (ESR) measurements were performed on NV centres within a nanopillar, both with and without a steady current of 1 A in the gradient microcoil. The magnetic field produced by the gradient microcoil induced a Zeeman shift of the ESR resonance line (in addition to that caused by the static magnetic field). A tem-perature change of the diamond sample due to current in the gradient microcoil also caused a decrease in the zero-field splitting of 74.2 kHz per degree rise in temperature (Ref. S2). From the observed 1 MHz shift in the zero-field splitting when a steady current of 1 A was sent though the microcoils, the temperature shift was estimated to be ≈13 K (Fig. S1(a)). In the second experimental technique, the change in resistance of the microcoil was measured as a function of current. From the known value of the temperature coefficient of resistance of gold, the temperature rise at 1 A current was estimated to be ≈13 K (Fig. S1(b)). The actual temperature rise of the diamond sample during Fourier magnetic imaging is expected to be less than these values due to the finite duty cycle of the gradient pulse sequence shown in Fig. 1d of the main text.

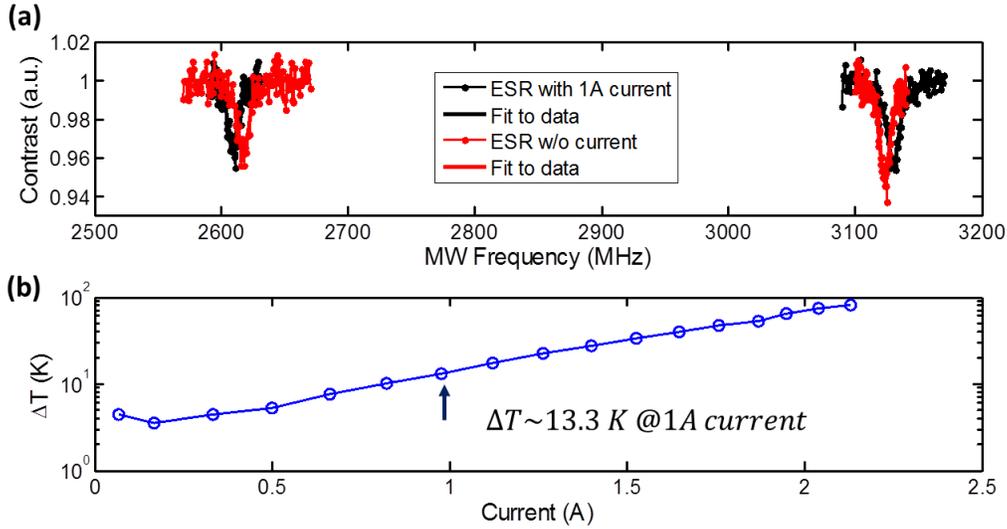

**Figure S1** | Measurements of diamond temperature change due to steady current through the gradient microcoil. (**a**) NV ESR spectral line shifts for 1 A current. The NV axial zero-field splitting parameter D varies with temperature T. By measuring the shift in D (midway between the $|0\rangle$ to $|\pm 1\rangle$ ESR frequencies), magnetic-field-induced (Zeeman) ESR line shifts can be removed and the diamond temperature change can be estimated. (**b**) Measurement of microcoil resistance change with increasing current. Here, the temperature coefficient of resistance for gold ≈0.0034 K$^{-1}$ is used. When 1 A current is applied, the estimated temperature increase ≈13 K, consistent with the result from ESR measurements.

### III. Gradient calibration

As described in the main text, NV signals for a point in k-space have the form $s(\vec{k}) \sim \cos(2\pi \vec{k} \cdot \vec{r}_0)$, where $\vec{k} = \gamma\tau(dB_\zeta/dx, dB_\zeta/dy)$. Thus the magnitude of k is proportional to the gradient, which is in turn proportional to the current I, and this current is proportional to the voltage V applied by the SRS345 programmable signal generator. The overall proportionality constant, which maps V to k, was estimated using the procedure described below.

A low-resolution one-dimensional phase encoding sequence was run sequentially on a set of 7 nanopillars along the X (and separately Y) direction. The NV signal was measured as a function of SRS345 voltage, V. The signal for each nanopillar was then Fourier transformed to obtain a peak in "inverse voltage" space. The peak positions (in units of Volt$^{-1}$) were then plotted as a function of nanopillar positions. The inter-pillar spacing of 1 micron, known from the defined lithographic pattern (see Methods), allowed conversion between Volt$^{-1}$ units and distance units in microns. This calibration procedure made accurate determination of local gradients possible and also properly accounted for any misalignment between the diamond coverslip patterned with microcoils and the NV-containing diamond sample. Calibration data for the Y micocoils is shown in Fig. S2 below. Similar results were obtained for the X coils.

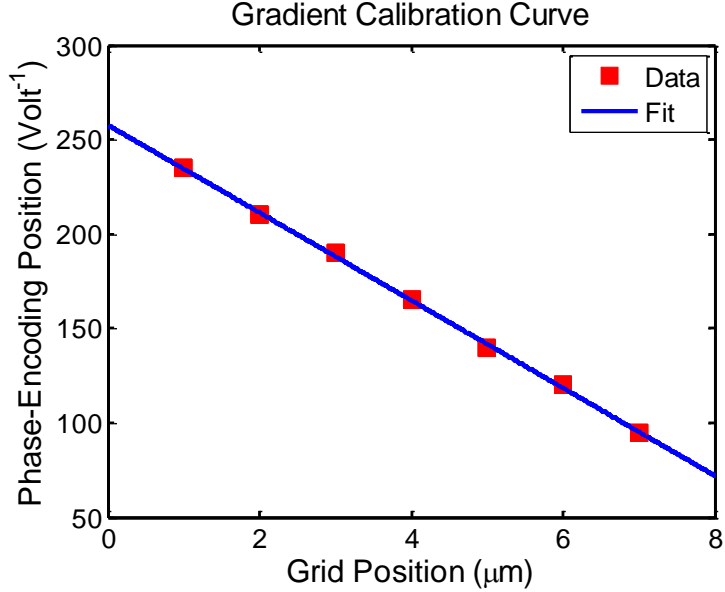

**Figure S2** | 1D low resolution phase encoding measurements for Y micocoil gradient calibration. The slope of the blue line is -23 V$^{-1}$/μm. The SRS345 voltage is divided by this value to obtain k-space data in μm$^{-1}$ units.

### IV. Magnetic field estimates and measured values for wide FOV image

For the three example nanopillar magnetic images shown in Fig. 4 of the main text, the measured AC magnetic field differences between NV centres are $\Delta B_{J=1} = [6.5 \pm 1.1] \times 10^2$ nT, $\Delta B_{J=2} = [1.9 \pm 0.6] \times 10^2$ nT, and $\Delta B_{J=3} = [0.8 \pm 0.4] \times 10^2$ nT. A measure of the long-range magnetic field gradient is provided by the low-resolution real-space image: $\vec{\nabla} B(\vec{r}_J) = [(B(x_{J+1}) - B(x_J))/\Delta x, \ (B(y_{J+1}) - B(y_J))/\Delta y]$, where $\vec{r}_J = [x_J, y_J]$ is the position of the J$^{th}$ nanopillar and $\Delta x, \Delta y = 1$ μm is the distance between adjacent nanopillars. The expected magnetic field difference between NV centres within the J$^{th}$ nanopillar is then given by: $\Delta B^{exp}(\vec{r}_J) = \vec{\nabla} B(\vec{r}_J) \cdot (\vec{r}_l - \vec{r}_m)$, where $\vec{r}_l, \vec{r}_m$ are the NV positions determined via Fourier imaging. The expected magnetic field differences for the three example nanopillars are: $\Delta B^{exp}_{J=1} = [5.9 \pm 1.4] \times 10^2$ nT, $\Delta B^{exp}_{J=2} = [2.0 \pm 1.1] \times 10^2$ nT, and $\Delta B^{exp}_{J=3} = [1.5 \pm 0.6] \times 10^2$ nT, which are in close agreement with measured values.

## V. Noise statistics and thresholding

The noise in k-space signals was observed to be Gaussian in nature. Upon Fourier transformation, the noise in the real and imaginary parts of the real-space signal was also observed to be Gaussian, as shown in Fig. S3. This is to be expected, as Fourier transformation is a linear operation and will not alter noise statistics. The noise in the absolute value of the real-space signal, however, was observed to follow a Rayleigh distribution (Fig. S3). This is well known in the theory of MRI (Ref. S3). The standard deviation of the Rayleigh distribution was computed and the real-space data was thresholded at 5 times this standard deviation to obtain the 2D images of NV centres shown in the main text.

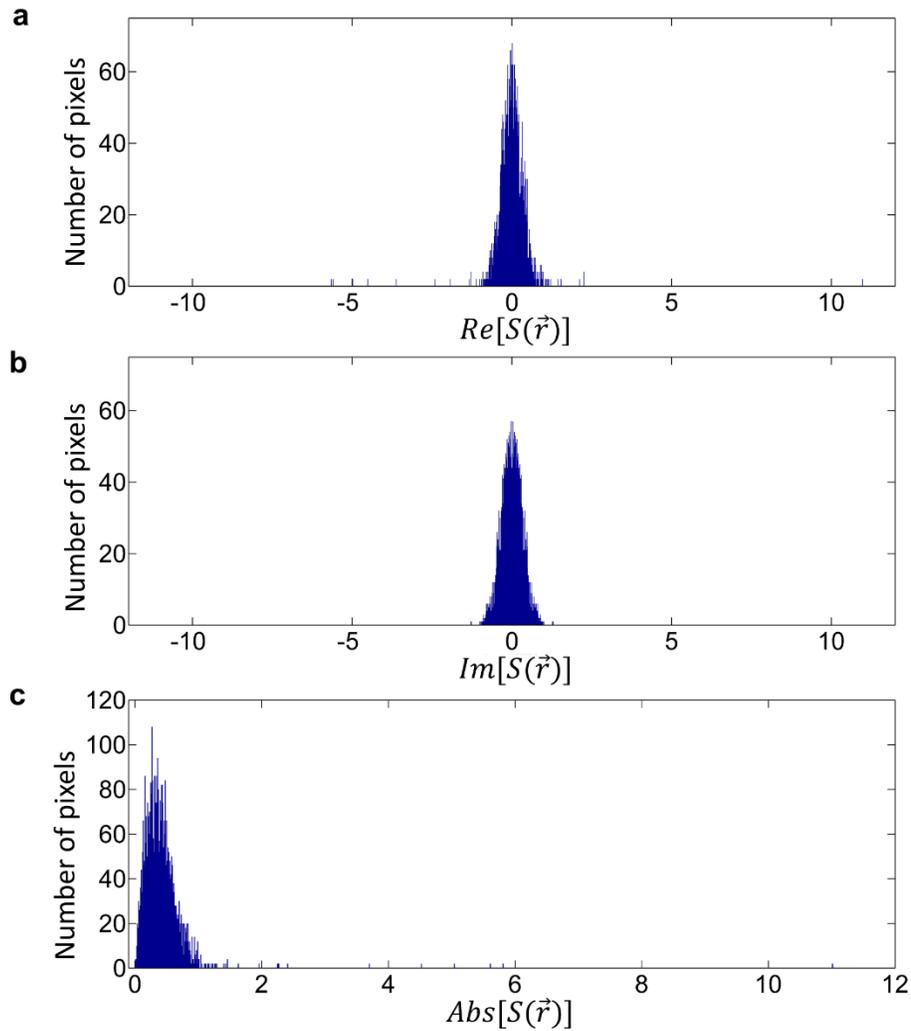

**Fig. S3** | Analysis of noise statistics. Fourier transform of k-space data from Fig. 2e of the main text. **(a)** Histogram of the real part of the Fourier transformed real-space signal. **(b)** Histogram of the imaginary part of the Fourier transformed real-space signal. The data of (a) and (b) exhibit Gaussian noise; and are fit to extract the standard deviation, σ of the noise. **(c)** Histogram of the absolute value of the Fourier transformed real-space signal. The noise follows a Rayleigh distribution.

## VI. Multiplex advantage

In this section we discuss the circumstances under which Fourier (k-space) imaging is faster than imaging based on point-by-point (real space) scanning. To make the discussion broad-based and not specific to NV centers, we make the following simplifying assumptions. We assume $N_E$ ideal photon emitters that emit $R$ photons per second, which are to be imaged using both point-by-point scanning and Fourier techniques. We further assume that each emitter is a two-level spin system with optical contrast =1; i.e., all $R$ photons are emitted by the spin-down state and zero photons by the spin-up state.

We first consider the case of point-by-point scanning. If the imaging device scans over $N_{pix}$ number of pixels spending time $dt$ per pixel, then the total imaging time is $N_{pix}dt$ and the SNR is $\sqrt{Rdt}$. The data acquisition time, defined as $T_{DAQ} = T/(N_{pix} * \text{SNR}^2)$, where $T$ is the total imaging time, is therefore given by $T_{DAQ} = 1/R$.

In Fourier imaging, data is first acquired in k-space. Assuming the number of k-space points is $N_{pix}$ and photons are counted for time $dt$ per pixel, the total imaging time is $N_{pix}dt$ and the SNR in k-space is $\sqrt{N_E Rdt}$. To compute SNR is real space (i.e., after Fourier transformation) we use Parseval's theorem and find $(Signal)_r = N_{pix}/\sqrt{N_E}(Signal)_k$ and $(Noise)_r = \sqrt{N_{pix}}(Noise)_k$, which implies $(\text{SNR})_r = \sqrt{N_{pix}/N_E}(\text{SNR})_k$. Here the subscripts $r$ and $k$ represent real space and k-space respectively. The SNR for Fourier imaging is therefore equal to $\sqrt{N_{pix}Rdt}$, which is higher than the SNR for point-by-point scannng by a factor $\sqrt{N_{pix}}$. Consequently, the total data acquisition time for Fourier imaging is $T_{DAQ} = 1/(R * N_{pix})$, which is less than that of point-by-point scanning by a factor $N_{pix}$. This improvement in SNR (or equivalently, multiplex speed-up in data acquisition time) is well known in FTIR spectroscopy, where it is referred to as the Fellgett advantage [S4].

Note that the full multiplex enhancement in SNR may not always be realized under realistic experimental conditions. For example, if the optical contrast is not perfect or if the point-spread-function of the signal peaks is wider than the imaging resolution, then the multiplex advantage will be diminished, though it may still be substantial. In Fig. S4 we use data from Figs. 2a and 2b of the main text to show how SNR in real space scales with the number of pixels under our experimental conditions. The SNR is observed to increase with the number of pixels albeit with a slightly smaller exponent of 0.36(8) instead of 0.5.

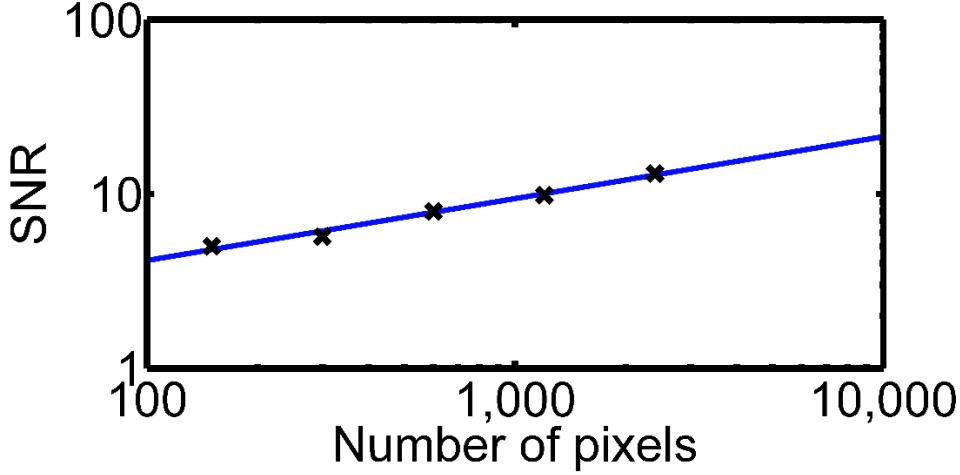

**Fig. S4** | Scaling of SNR with number of number of pixels. Fit to data (blue line) shows SNR scales as $N_{pix}^{0.36(0.8)}$ which is close to but smaller than the expected $N_{pix}^{0.5}$ scaling.

## VII. Gradient sensitivity

The gradient sensitivity between the two NV centers in Fig. 3 of the main text is calculated from the uncertainty of the field gradient, which is given by $G = \Delta B/\Delta x$, where $\Delta B = B(x_1) - B(x_2) = 650$ nT is the measured field difference between the two NV centers and $\Delta x = x_1 - x_2 = 121$ nm is the distance between them. Through error propagation the minimum detectable gradient is found to be $\delta G = G\,[(\delta \Delta B/\Delta B)^2 + (\delta \Delta x/\Delta x)^2]^{1/2}$ where $\delta \Delta B = \sqrt{2}\ast \delta B = \sqrt{2}\ast(1200$ nT $/\sqrt{T})$ is the uncertainty in the magnetic field difference, and $\delta \Delta x = \sqrt{2}\ast \delta x \sim \sqrt{2}\ast (30$nm $/$SNR$/\sqrt{T})$ is the uncertainty in the NV separation. Here T represents the total measurement time. Note that the dominant contribution comes from the first term. The magnetic field gradient sensitivity is thus given by $\eta_G = \delta G\sqrt{T} = (650$ nT$/121$nm$)[(\sqrt{2}\ast 1200$ nT$/650$nT$)^2 + (\sqrt{2}\ast 30$nm$/$SNR$/110$nm$)^2]^{1/2} = 14$ nT/nm/$\sqrt{Hz}$.

## VIII. Nanopillar map

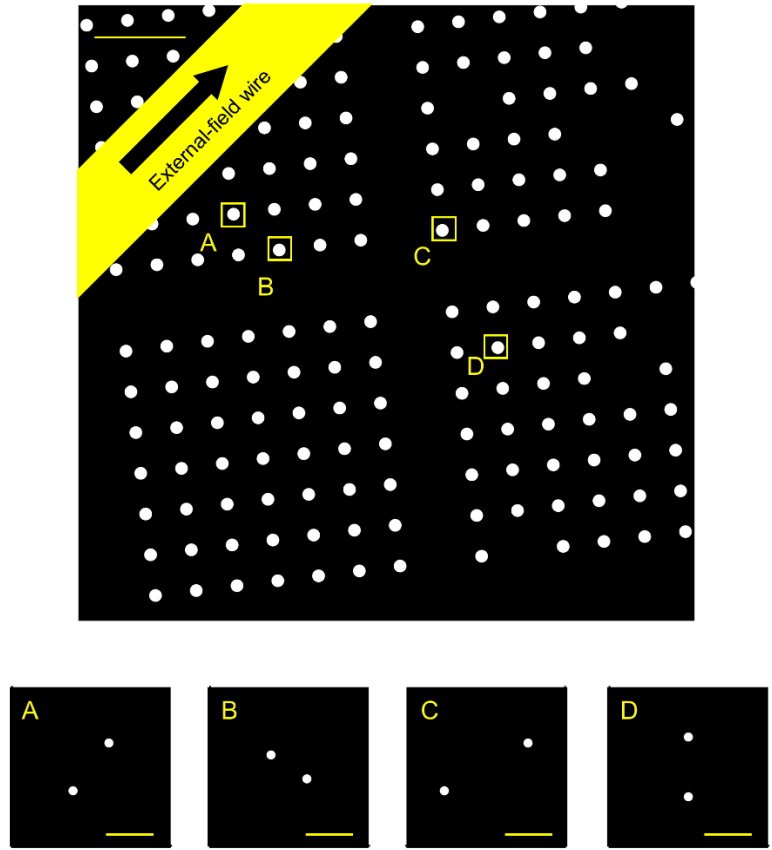

**Fig. S5** | Location of nanopillars relative to the external-field wire. Data for Figs. 2c-f and Fig. 3 of the main text were taken from nanopillar B. Data for Fig. 4 of the main text was taken from nanopillars B, C, and D. The compressed sensing data of Fig. 5 of the main text was taken from nanopillar A. Scale bars are 2 μm and 100 nm for top and bottom figures, respectively.